\documentclass[pra,twocolumn,showpacs,a4paper]{revtex4}
\usepackage[latin1]{inputenc}
\usepackage{amsmath,amssymb}
\usepackage{mathrsfs}
\usepackage{graphicx}
\usepackage[T1]{fontenc}

\newcommand{\ave}[1]{\ensuremath{\left \langle {#1} \right \rangle}}
\newcommand{\abs}[1]{\ensuremath{\left\lvert{#1}\right\rvert}}

\newcommand{\spectral}{\ensuremath{\mathscr{S}}}
\newcommand{\ii}{\ensuremath{\mathfrak{i}}} 
\newcommand{\jj}{\ensuremath{\mathfrak{j}}}

\newcommand{\al}{\ensuremath{\alpha}}

\newcommand{\de}{\ensuremath{\delta}}
\newcommand{\De}{\ensuremath{\Delta}}
\newcommand{\om}{\ensuremath{\omega}}
\newcommand{\Om}{\ensuremath{\Omega}}

\newcommand{\E}{\ensuremath{\mathcal{E}}}

\newcommand{\V}{\ensuremath{\mathcal{V}}}

\begin{document}


\title{Sub-Poissonian laser emission from a single-electron permanently interacting with a single-mode cavity.}

\author{Jacques Arnaud}

\email{arnaudj2@wanadoo.fr}

\affiliation{Mas Liron, F30440 Saint Martial, France}

\author{Laurent Chusseau}

\email{chusseau@univ-montp2.fr}

\affiliation{Institut d'Électronique du Sud, UMR 5214 CNRS, Université Montpellier II, F34095
Montpellier, France}

\author{Fabrice Philippe}
\altaffiliation[Also at ]{MIAp, Université Paul Valéry, F34199 
Montpellier, France}

\email{philippe@lirmm.fr}

\affiliation{Laboratoire d'Informatique de Robotique et de
Microélectronique de Montpellier, UMR 5506 CNRS, 161 Rue Ada, F34392
Montpellier, France}

\date{\today}

\begin{abstract}      
Quiet (or sub-Poissonian) oscillators generate a number of dissipation events whose variance is less than the mean. It was shown in 1984 by Golubev and Sokolov that lasers driven by regular pumps are quiet in that sense. The purpose of this paper is to show that, as long as the laser-detector system is strictly stationary, quantization of the optical field is not required to explain such phenomena. The theory presented here is semi-classical, yet exact. Previous theories considering excited-state atoms regularly-injected in resonators, on the other hand, do require in principle light quantization. Specifically, we consider a laser involving a single electron permanently interacting with the field and driven by a constant-potential battery, and point out a similarity with reflex klystrons. The detected noise is found to be only 7/8 of the shot-noise level. It is therefore sub-Poissonian. Our calculations are related to resonance-fluorescence treatments but with different physical interpretations.
\end{abstract}

\pacs{42.55.Ah, 42.50.Ar, 42.55.Px, 42.50.Lc}

\maketitle

\section{Introduction}\label{introduction}

In many experiments, we only need to know time-averaged photo-currents. This information suffices for example to verify that light passing through an opaque plate pierced with two holes exhibits interference patterns. The experiment is performed by measuring the time-averaged photo-currents issued from an array of detectors located behind the plate. On the other hand, experiments involving the transmission of information through an optical fiber require that the fluctuations of the photo-current about its mean be known \cite{note1}. The information to be transmitted is corrupted by natural fluctuations (sometimes referred to as "quantum noise"). Laser noise impairs the operation of optical communication systems and the measurement of small displacements or small rotation rates with the help of optical interferometry. Even though laser light is far superior to thermal light, minute fluctuations restrict the ultimate performances. Signal-to-noise ratios, displacement sensitivities, and so on, depend mainly of the spectral densities, or correlations, of the photo-electron events. It is therefore important to have at our disposal formulas enabling us to evaluate these quantities for configurations of practical interest, in a form as simple as possible. We are mostly concerned with basic concepts leaving out detailed practical calculations. Real lasers involve many secondary effects that are neglected here for the sake of clarity.

A quiet oscillator generates a number of dissipation events whose variance is less than the mean. Equivalently, when the photo-current $j(t)$ is analyzed in the Fourier domain, the (double-sided) spectral density of the photo-current is smaller than the product of electron charge $e$ and average current $\ave{j}$ as the angular frequency $\Om\to 0$ (sub-Poissonian light). It was shown in 1984 by Golubev and Sokolov \cite{golubev:JETP84} that lasers driven by regulated pumps are quiet in that sense. 

We consider in the present paper a battery-driven laser involving a single electron permanently interacting with the field, and point out a similarity with reflex klystrons. The theory presented does not require field quantization and is therefore "semi-classical", yet exact except for the approximation made in every above-threshold laser theory that the fluctuations considered are small and slow and the power transferred from the static source to the optical load is small. We find that for a one-electron laser driven by a constant-potential battery the detected noise is 7/8 of the shot-noise level and is therefore sub-Poissonian. Our calculations are related to resonance-fluorescence treatments but they have a different physical interpretation. In contradistinction, previous theories considering instead excited-state atoms regularly injected in resonators do require in principle light quantization. For a review of important theoretical and experimental papers on that subject see the collection in \cite{Meystre1991a}. 

The present semi-classical theory \cite{Arnaud1990a} is accurate and easy to apply. Once the necessary assumptions have been agreed upon, laser noise formulas for various configurations follow from elementary mathematics. In particular, operator algebra is not needed. For simple laser models, e.g.~incoherently-pumped 4-level lasers \cite{Chusseau2002b}, there is exact agreement between our results and those derived from quantum optics for any parameter values. Quantization of the field is clearly required when an atom in the upper state is injected into an empty resonator. Provided that the atom transit time has some well-defined value, the exiting atom is with certainty in the lower state, a conclusion that, to our knowledge, cannot be explained without quantizing light. Theories considering excited-state atoms regularly-injected in low-loss resonators require in principle light quantization, although some approximation may reduce them to simpler rate equations in the high-field limit. In the present configuration the electron interacts permanently with the field in a strictly stationary manner thus allowing the field quantization to be ruled out.

The semi-classical theories employed in optical engineering, on the other hand, rest on the concept that the classical oscillating field is supplemented by a random field due to spontaneous emission. Such semi-classical theories are unable, however, to describe sub-Poissonian light, and are therefore to be distinguished from the present theory. From our view-point, spontaneous emission is unessential, and is neglected for simplicity. The first sections report well-known results, namely the Rabi-oscillation theory, see e.g.~\cite{Scully1997}. These sections should enable the reader to follow the paper throughout starting from elementary classical considerations. Results obtained from the present theory for configurations of practical interest were listed in \cite{Arnaud06}.

\section{Configuration}\label{klystronsec}

The laser that we consider is very similar to a battery-driven reflex klystron, see e.g.~\cite{Arnaud1990}, except for the fact that we suppose that a single electron interacts with the resonator. The only difference existing between a microwave oscillator such as a reflex klystron and a laser relates to the different electronic responses to alternating fields. In a microwave tube the electron natural motion is usually not harmonic and its coupling to single-frequency electromagnetic fields may be understood accurately only through numerical calculations. In contradistinction, masers and lasers employ basically two-level molecules or atoms, and this results in simplified treatments. The phenomenon of stimulated emission is essentially the same for every oscillator. Let us quote the Nobel-prize winner W. E. Lamb, Jr.~\cite[p.~208]{Lamb2001}: "Whether a charge $q$ moving with velocity $dx/dt$ in an electrical field $\E$ will gain or loose energy depends on the algebraic sign of the product $q\E dx/dt$. If the charge is loosing energy, this is equivalent to stimulated emission". 

\begin{figure}
\centering
\includegraphics[width=0.9\columnwidth]{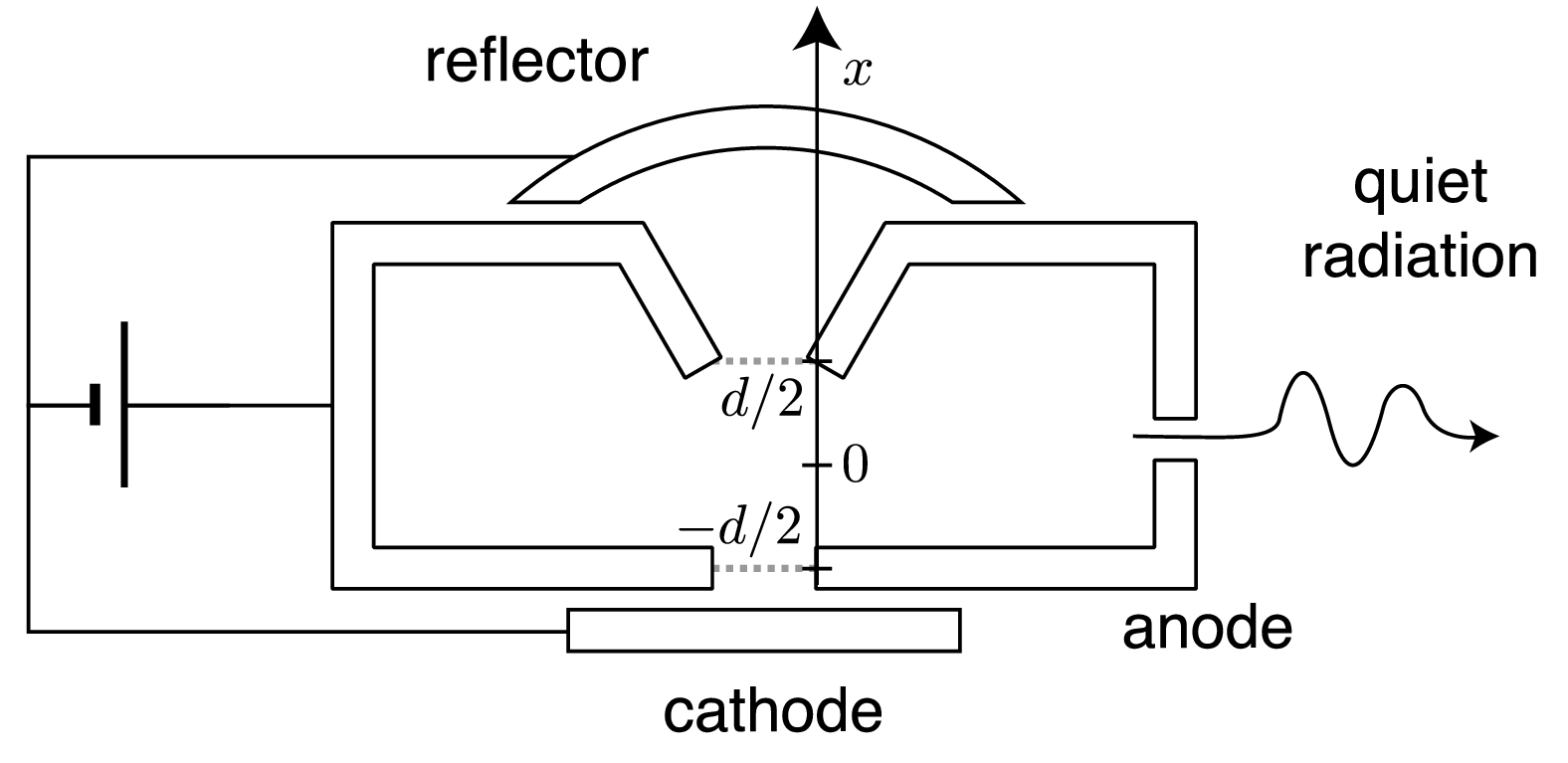} 
\caption{ The laser considered is analogous to the reflex klystron represented in this figure, but involves a single electron. This electron moves essentially along the vertical $x$-axis, being guided by an $x$-directed magnetic field (not shown). When this electron has lost most of its energy it may be captured by the anode, being replaced by a high-energy electron. The inner part of the resonator is modeled as a capacitance $C$ with grids spaced a distance $d$ apart and the outer part by an inductance $L$, with resonant angular frequency $\om$. In the laser version the electron may reside in only two states, labeled 1 and 2, with energy separation equal to $\hbar\om$. The separations between the upper grid and the reflector and between the lower grid and the cathode are assumed to be negligibly small, so that electrons are instantly reflected.}
\label{klystron}
\end{figure}

A reflex klystron is schematized in Fig.~\ref{klystron}. Consider an electron located between two parallel conducting plates pierced with holes (or with grids) called "anodes", and constrained to move essentially along the vertical $x$-axis. A static potential source $U$ is applied to external plates (cathode and reflector) to reflect the electron. There is an alternating potential $v(t)=v\cos (\om t)$ between the two anodes (grids) when the klystron oscillates. Once an electron has lost most of its energy, it moves side-ways, gets captured by anodes, and an electron is emitted back by the cathode. The over-all effect of the electron motion is therefore to transfer energy from the static potential $U$ to the alternating potential $v(t)$, an effect analogous to stimulated emission. In the classical treatment, one first evaluate 1) the electron motion under the static field, 2) the perturbation caused by alternating fields, called "bunching", and 3) the current induced in the alternating potential source. The same steps are taken in the quantum treatment. Namely, we consider the stationary states of an electron submitted to a static field, the perturbation of those states due to alternating fields, and finally evaluate the current induced by the electron motion.

The static field between the anodes ($-d/2<x<d/2$) vanishes, but the electron is reflected by the cathode and reflector. Because of the applied alternating potential $v(t)=v\cos(\om t)$, the electron is submitted to a potential $v\cos(\om t)x/d$ when $-d/2<x<d/2$. The classical equations of motion of an electron of charge $-e$, mass $m$ are best based on the Hamiltonian formulation in which the particle energy $E(t)$ is expressed as a function of position $x$, momentum $p$, and time $t$, according to the relation
\begin{align}
\label{ham}
H(x,p,t)-E(t)\equiv \frac{p^2}{2m}-ev\cos(\om t)\frac{x}{d}-E(t)=0,
\end{align}
where $p^2/(2m)$ represents the kinetic energy. The Hamiltonian equations read
\begin{align}
\label{eqmot}
\frac{dx(t)}{dt}&=\frac{\partial H(x,p,t)}{\partial p }=\frac{p(t)}{m}\\
\frac{dp(t)}{dt}&=-\frac{\partial H(x,p,t)}{\partial x}=\frac{ev}{d}\cos(\om t).
\end{align}
The first equation says that the particle momentum $p(t)=m dx(t)/dt$, and the second equation may be written, with the help of the first equation, in the usual Newtonian form (force=mass $\times$ acceleration).

To evaluate the induced current consider again two conducting plates a distance $d$ apart submitted to a potential source $v(t)$, and an electron of charge $e$ in between. If $i(t)$ denotes the current delivered by the potential source, the power $v(t)i(t)$ delivered by the source must be equal at any instant to the power received by the electron, which is the product of velocity $dx(t)/dt$ and force $ev(t)/d$, that is, $v(t)i(t)=e\frac{v(t)}{d}\frac{dx(t)}{dt}$. Since $v(t)$ drops out, the current induced by the electron motion is
\begin{align}
\label{currentb}
i(t)=\frac{e}{d}~\frac{dx(t)}{dt}.
\end{align}
When the alternating potential $v(t)$ depends on the delivered current $i(t)$ the full circuit equations must be solved. 

\section{The Schrödinger equation}\label{lasers}

The lasers considered oscillate in a single electromagnetic mode in the steady state. Only the stationary regime is treated, that is, the system elements do not depend explicitly on time and fluctuation correlations are independent of the initial time. 

We consider the same configuration as in Fig.~\ref{klystron} with an electron constrained to move along the $x$-axis. Its motion is described by a wave-function $\psi(x,t)$ satisfying the Schrödinger equation
\begin{align}
\label{sch}
\big[ H(x,p,t)-E \big] \psi(x,t)=0,
\end{align}
where $E=\ii \hbar\partial/ \partial t$, $p=-\ii \hbar\partial/ \partial x$ and $H(x,p,t)=\frac{p^2}{2m} - e  v \cos(\om t)\frac{x}{d}$ as in the previous section, but $p$ and $E$ are now operators of derivation. It is easily shown that, provided $\psi(x,t)$ decreases sufficiently fast as $x\to±\infty$, the integral over all space of $\abs{\psi(x,t)}^2$ does not depend on time, and therefore remains equal to 1 if the initial value is 1, a result consistent with the Born interpretation of the wave function. Because of linearity the sum of solutions of the Schrödinger equation is a solution of the Schrödinger equation (superposition state). The cathodes in Fig.~\ref{klystron} (electron-emitting cathode and reflector) reflect quickly the electron back to the interaction region. This is expressed by specifying that the wavefunction $\psi(x,t)$ vanishes when $\abs{x} \ge d/2$ .

The above Schrödinger equation enables us to evaluate the motion of an electron submitted to a deterministic potential $v(t)$. The induced current is related to the electron velocity as in the classical case, see \eqref{currentb}, except that $i$ and $x$ are being replaced by their quantum-mechanical expectation values
\begin{equation}
\label{current2}
\begin{aligned}
\ave{i(t)}&=\frac{e}{d}\frac{d\ave{x(t)}}{dt}\\
\ave{x(t)}&\equiv \int_{-d/2}^{d/2}dx~x~\abs{\psi(x,t)}^2.
\end{aligned}
\end{equation}
The average power received by the deterministic alternating potential $v(t)$ reads $\ave{P(t)}=v(t)\ave{i(t)}$, and the average energy received from time $t=0$ to $\tau$ is the integral from 0 to $\tau$ of $v(t)\ave{i(t)}$.

\section{Static potential}\label{static}

When $v(t)=0$ the Schrödinger equation \eqref{sch} admits solutions of the form $\psi_n(x,t)=\psi_n(x)\exp(-\ii\om_nt)$, where $n=1,2$. With the boundary conditions specified above, namely $\psi_n(±d/2)=0$, the lowest-energy state $n=1$ and the first excited state $n=2$ are described by the wave-functions
\begin{equation}
\begin{aligned}
\label{wf}
\psi_{1}(x,t)&=\sqrt{2/d} \cos(\pi x/d)\exp(-\ii \om_1 t)\\
&\equiv\psi_{1}(x)\exp(-\ii \om_1 t)\\
\psi_{2}(x,t)&=\sqrt{2/d} \sin (2\pi x/d)\exp(-\ii \om_2 t)\\
&\equiv\psi_{2}(x)\exp(-\ii \om_2 t),
\end{aligned}
\end{equation}
that satisfy the usual orthonormality condition. Energy eigenvalues are straightforwardly obtained by substituting \eqref{wf} in \eqref{sch}
\begin{align}\label{En}
E_{n}\equiv \hbar \om_n=\frac{\pi^2 \hbar^2}{2md^2}n^2, \quad \text{with} \quad n=1,2.
\end{align} 
We will see later on that optical fields at frequency
\begin{align}\label{om}
\om=\om_2-\om_1=\frac{3\pi^2\hbar}{2md^2}
\end{align} 
may cause the system to evolve from state 1 to state 2 and back. For illustration, let us select a frequency $\nu\equiv\om/2\pi$=1.42~GHz (hydrogen hyperfine-transition frequency). Then \eqref{om} gives $d=0.44\;\mu$m.

For later use let us evaluate
\begin{align}\label{x12}
x_{12}\equiv \int_{-d/2}^{d/2}dx~x~\psi_{1}(x)\psi_{2}(x)=\frac{16d}{9\pi^2}.
\end{align} 
The parameter $x_{12}$ determines the strength of the atom-field coupling through the Rabi frequency defined by
\begin{align}
\label{rabi}
\hbar\Om_R \equiv \frac{evx_{12}}{d}=\frac{16}{ 9\pi^2}ev,
\end{align}
where $v/d$ denotes the peak applied field (see the next section). Let the anodes represent a capacitance $C=\epsilon_o A/d$ (where $\epsilon_o$ denotes the free-space permittivity, $d$ the spacing, and $\V=Ad$ the capacitance volume) connected to an inductance $L$ such that $LC\om^2=1$, where the angular optical frequency $\omega$ was defined in \eqref{om}. The classical expression of the average resonator energy is $E=Cv^2/2$. From \eqref{rabi} and the above relations, the square of the Rabi frequency may be written as 
\begin{equation}
\label{E}
\Om_R^2=b\frac{\mu}{\V}
\end{equation}
with
\begin{equation*}
\mu\equiv \frac{E}{\hbar\om} \quad \text{and} \quad 
b\equiv \frac{1024}{27\pi}\frac{e^2}{4\pi\epsilon_o~ m}\approx 3000~ \textrm{m}^3/\textrm{s}^2
\end{equation*}
showing that $\Om_R^2$ is proportional to the resonator energy.

\section{Perturbed motion}\label{perturbed}

We next suppose that a potential source $v(t)=v\cos(\om t)$ is applied between the two anodes in Fig.~\ref{klystron}, a distance $d$ apart. The electron is submitted to a potential $-ev\cos(\om t)x/d$ where $\om$ is the 1-2 transition frequency defined in \eqref{om}. In that case \eqref{sch} reads         
\begin{align}
\label{solve}
\left(\frac{p^2}{2m}-\frac{evx}{d}\cos(\om t)-E\right)\psi(x,t)=0.
\end{align}
Supposing that, as a result of the resonance condition, only states 1 and 2 are significant, the wave-function may be written as 
\begin{multline}
\label{solveter}
\psi(x,t)=C_1(t)\exp(-\ii \om_1 t)\psi_1(x)\\
+C_2(t)\exp(-\ii \om_2 t)\psi_2(x)
\end{multline}
with slowly time-varying coefficients $C_1(t),C_2(t)$. If we substitute this expression into the Schrödinger equation \eqref{solve} and take \eqref{wf} into account we obtain 
\begin{multline}
\label{solebis5}
0=\sum_{n=1}^2\exp(-\ii \om_n t)\\
\cdot\left(\ii\hbar\frac{dC_n(t)}{dt}+\frac{evx}{d}\cos(\om t)~C_n(t)\right)\psi_n(x).
\end{multline}
Multiplying \eqref{solebis5} throughout by $\psi_m(x), ~m=1,2$, integrating with respect to $x$, and taking into account the orthonormality of the $\psi_n(x)$ functions, we obtain a pair of differential equations for $n=1,2$
\begin{equation}
\label{lveter}
0=\ii\hbar\frac{dC_n(t)}{dt}+\exp(-\ii \om t)\cos(\om t)\frac{evx_{12}}{d}C_n(t).
\end{equation}
The rotating-wave approximation consists of keeping only slowly-varying terms, that is, replacing $\exp(-\ii \om t)\cos(\om t)$ by 1/2. Thus, the complex coefficients $C_1(t),~C_2(t)$ obey the differential equations
\begin{equation}
\label{formbis}
\begin{aligned}
\frac{dC_1(t)}{dt}&=\ii\frac{\Om_R}{2}C_2(t)\\
 \frac{dC_2(t)}{dt}&=\ii\frac{\Om_R}{2}C_1(t), 
\end{aligned}
\end{equation}
where $\Om_R$ is the Rabi frequency defined earlier and the orthonormality reads $C_1(t)C_1^\star(t)+C_2(t)C_2^\star(t)=1$. These differential equations are easily solved. Assuming that the electron is initially ($t=0$) in the upper state, the probability that the electron be found in the lower state at time $t$ reads $C_1(t)C_1^\star(t)=\sin^2(\Om_R t/2)$. Initially, the electron delivers energy to the alternating potential (stimulated emission) but this energy is subsequently recovered by the electron, so that no laser action may be achieved. This is the usual Rabi solution for two-level atoms at resonance.

\section{Electron jump through a battery}\label{damped}

We suppose that the laser has reached a steady-state of oscillation so that the electron is submitted to an alternating potential at frequency $\om$ as discussed above. The electron, initially in the upper-energy state 2 slowly acquires a non-zero probability of being in the lower-energy state 1, thereby delivering energy to the alternating potential. This is the phenomenon of stimulated \emph{emission} of radiation. 

But we now postulate that this process may be interrupted by the fact that an electron in the lower-energy state 1 may jump instantly to state 2 through the battery, as shown in  Fig.~\ref{klystron}, discharging it slightly, through some kind of tunneling effect. This is how the electron may recover the energy delivered to the alternating potential, at least on the average. The probability density of a jump from state 1 to state 2 through the battery will be denoted $2\gamma$. The average time between successive jumps will be denoted by $\ave{\tau}$. This is a function of $\gamma$ and of the alternating potential strength (or Rabi frequency) given later on in \eqref{avt}. Note that if the electron is in state 2 at time $t=0$, as we later assume, this implies that a jump from state 1 to state 2 just occurred. If $G(t)dt$ denotes the probability that another jump (not necessarily the next one) occurs between $t$ and $t+dt$, $G(t)$ is the correlation between jumps separated in time by $t$ (remembering that the jump process considered is stationary). This correlation will be obtained by evaluated first the probability density $w(t)$ that the \emph{next} jump occurs at time $t$.

The equations in \eqref{formbis} generalize to
\begin{equation}
\label{formx}
\begin{aligned}
\frac{dC_1(t)}{dt}&=\ii\frac{\Om_R}{2}C_2(t)-\gamma C_1(t)\\
\frac{dC_2(t)}{dt}&=\ii\frac{\Om_R}{2}C_1(t).
\end{aligned}
\end{equation}
The equation obeyed by $C_1(t)$ is obtained by deriving the first equation with respect to time and employing the second equation. We obtain
\begin{align}
\label{formx2}
 \frac{d^2C_1(t)}{dt^2}+\gamma \frac{dC_1(t)}{dt}+ \left( \frac{\Om_R}{2} \right)^2 C_1(t)=0.
\end{align}
When the electron is initially in the upper-energy state 2, the initial condition is $C_2(0)=1,~C_1(0)=0$, and we find from \eqref{formx2}
\begin{equation}
\label{formy}
C_1(t)=
\frac{\ii\Om_R}{2\al}\left(\exp\left(\frac{\al-\gamma}{2}t\right)-\exp\left(\frac{-\al-\gamma}{2}t\right)\right)
\end{equation}
with $ \al\equiv\sqrt{\gamma^2-\Om_R^2}$.

The quantity $C_1(t)C_1^\star(t)$ represents the probability that the electron resides in the lower state \emph{as long as no jump occurs}. This quantity tends to decay in the course of time according to the $\exp(-\gamma t)$ factor, because, if no jump has occurred up to time $t$, most likely the electron does not reside in the lower state. 

We obtain directly from \eqref{formy} the waiting-time probability density
\begin{multline}
\label{wait}
w(t)=2\gamma C_1(t)C_1^\star(t)=\frac{\gamma\Om_R^2}{2\al^2}\Big( \exp ((\al-\gamma) t)\\
+ \exp(-(\al+\gamma) t) -2\exp(-\gamma t)\Big) .
\end{multline}
This expression (valid for any non-negative value of $\gamma$) was obtained before in connection with resonance fluorescence in \cite{Carmichael1989} through a different method. The quantity $w(t)dt$ is the probability that, given that the electron is in the upper state at $t=0$, it performs a jump from state 1 to state 2 \emph{for the first time} between $t$ and $t+dt$. When such a jump occurs, the same process starts again. Thus the jumps form an ordinary renewal process. The average inter-event time
\begin{align}
\label{avt}
\ave{\tau}=\frac{1}{R}\equiv\int_0^\infty dt~t~w(t)=\frac{1+2\gamma^2/\Om_R^2}{\gamma}\equiv\frac{1+a}{\gamma}
\end{align}
where $a\equiv2\gamma^2/\Om_R^2$ and $R$ the average jump rate.

The Laplace transform $\tilde{w}(p)$ of $w(t)$ in \eqref{wait} reads
\begin{align}
\label{rhomp}
\tilde{w}(p)=\frac{\gamma\Om_R^2}{p^3+3\gamma p^2+(2\gamma^2+\Om_R^2)p+\gamma\Om_R^2}.
\end{align}

It is straightforward to go from the waiting time probability density $w(t)$ evaluated above to the event probability density $G(t)$. The concept is that the probability density of an event occurring at $t$ is the sum of the probabilities that this occurs through 1 jump, 2 jumps,\dots\ It follows that $G(t)=w(t)+w(t)*w(t)+w(t)*w(t)*w(t)+\dots$, where the middle stars denote convolution products. Thus the Laplace transform $\tilde{G}(p)$ of $G(t)$ is the sum of an infinite geometric series, which may be written in terms of the Laplace transform $\tilde{w}(p)$ of $w(t)$ as \cite{Cox1980}          
\begin{align}
\label{rhoml}
\tilde{G}(p)=\frac{\tilde{w}(p)}{1-\tilde{w}(p)}.
\end{align}

The jump rate may be written in general as $R(t)=R+r(t)$ where $r(t)$ represents a small fluctuation. The quantity we are interested in is the (double-sided) spectral density $\spectral_{R(t)}(\Om)$ of the jumps at Fourier (angular) frequency $\Om$. According to the Wiener-Khintchine theorem, the spectral density is the Fourier transform of the event correlation, see for example \cite[\S 3.13]{Arnaud06}. It may thus be obtained directly from $\tilde{G}(p)$ after some rearranging as
\begin{align}
\label{oml}
\frac{\spectral_{R(t)}(\Om)}{R}&=\lim_{\epsilon\to 0} \left(1+\tilde{G}(\epsilon+\jj\Om)+\tilde{G}(\epsilon-\jj\Om)\right) \nonumber\\
&\equiv 2 \pi R \de(\Om)+\frac{\spectral_r(\Om)}{R}
\end{align}
\begin{multline*}
\text{with} \quad \frac{\spectral_r(\Om)}{R}=1\\
-\frac{3a}{(1+a)^2+a\left((5a/4)-1\right)\left(\Om/\gamma\right)^2+(a^2/4)\left(\Om/\gamma\right)^4}.
\end{multline*}
The term $2\pi R\de(\Om)$ simply expresses that the average rate equals $R$. The subsequent terms may be obtained by subtracting $R/p$ from $\tilde{G}(p)$, setting $p=\jj\Om$, and rearranging. We consider particularly the $\Om\to 0$ limit of $\spectral_r(\Om)$
\begin{equation}
\label{ol}
\frac{\spectral_r(0)}{R}=1-\frac{3a}{(1+a)^2}.
\end{equation}
In the large-$\gamma$ limit the shot-noise level $\spectral_r(0)=R$ is recovered. 

In the case of stimulated absorption, the role of the upper and lower states should be interchanged. Stimulated absorption occurs in optical detectors. Then the jumps previously considered correspond to photo-electron emission events. Unless the optical power is very large, detectors are linear. This weak-field condition corresponds in previous expressions to the case where $\Om_R\ll \gamma$. The fluctuations are then seen to be at the shot-noise level. Alternatively, one may employ in that limiting situation Nyquist-like noise sources \cite{Arnaud1990a}.

\section{Steady-state}\label{steady}

Going back to the configuration represented in Fig.~\ref{klystron}, note that the battery represented on the left delivers a measurable average electron rate $J$ that may be increased by increasing the battery potential $U$ slightly above $\hbar\om/e$ \cite{note2}. The rate $R$ generated by the electron is given in \eqref{avt}. Finally, radiation escaping from the hole shown on the right of the resonator is eventually absorbed by an ideal detector at a rate $D=\mu/\tau_p$, where the lifetime $\tau_p$ depends on the hole size. Evaluating $\tau_p$ is a classical electromagnetic problem that we assume solved. Thus, the steady state condition $J=R=D$ reads explicitly 
\begin{align}
\label{dy}
J=\frac{\gamma}{1+2\gamma^2/\Om_R^2}=\frac{\mu}{\tau_p}.
\end{align}
Accordingly, given the average electron-injection rate $J$ and the resonator lifetime $\tau_p$, we may evaluate the reduced resonator energy $\mu\equiv E/\hbar\om=J\tau_p$. Next, given the capacitance volume $\V$, we may evaluate $\Om_R^2$ from \eqref{E}, and the decay constant $2\gamma$ from \eqref{dy}. This value of $\gamma$ corresponds to some value of the static potential $U$, slightly above $\hbar\om/e$.

\section{Laser noise}\label{noise}

What we call "laser noise" refers to photo-current fluctuations. The result given in \eqref{ol} provides the rate-fluctuation spectral density for an electron submitted to an alternating potential independent of the electron motion. But in lasers the reduced resonator energy $\mu(t)=\mu+\De \mu(t)$ fluctuates. Because this fluctuation is small in above-threshold lasers the fluctuation $r(t)$ previously evaluated is supposed to be unaffected. The rate equation is
\begin{equation}
\label{rat2}
\frac{d\mu(t)}{dt}=R(t)-D(t),
\end{equation}
with
\begin{equation*}
R(t)=R+\frac{dR}{d\mu}\De \mu(t)+r(t), 
D(t)=\frac{\mu(t)}{\tau_p}+d(t),
\end{equation*}
and where $d\mu(t)/dt$ represents the rate of increase of the reduced resonator energy. This is the difference between the in-going rate $R(t)$ and the out-going (or detected) rate $D(t)$. Note that the in-going rate involves a term expressing the fact that $R$, as given in \eqref{avt}, depends on $\mu$ and that $\mu$ is now allowed to fluctuate. The outgoing rate is fully absorbed by an ideal cold detector at an average rate $D$, supplemented by a fluctuating rate $d(t)$, whose spectral density is equal to the average rate $D=R=J$. Because the noise sources $d(t)$ and $r(t)$ have different origins they are independent. 

Considering only fluctuating terms at zero Fourier frequency ($d/dt\to0$), we obtain $\De R(t)=\De D(t)$, that is, explicitly
\begin{align}
\label{rat}
\frac{dR}{d\mu}\De \mu(t)+r(t)=\frac{\De \mu(t)}{\tau_p}+d(t).
\end{align}
Solving this equation first for $\De \mu(t)$, with $\tau_p=\mu/R$, and substituting the result in the expression for $\De D(t)$, we obtain
\begin{align}
\label{rat3}
\De D(t)&\equiv \frac{\De \mu(t)}{\tau_p}+d(t)=\frac{r(t)-Ad(t)}{1-A} \\
\intertext{with}
A&\equiv \frac{\mu}{R} \frac{dR}{d\mu}=\frac{a}{1+a},
\end{align}
according to \eqref{avt} and \eqref{E}. Because $r(t)$ and $d(t)$ are independent, the spectral density of the photo-detection rate is, with $\spectral_r/D=1-3a/(1+a)^2$ from \eqref{ol} and $\spectral_d/D=1$,
\begin{align}
\label{rat5}
\spectral_{\De D}=\frac{\spectral_r+A^2 \spectral_d}{(1-A)^2}=(2a^2-a+1)D.
\end{align}
The smallest detector noise, obtained when $a=1/4$, is 7/8 of the shot-noise level. Therefore, a sub-Poissonian laser may be realized with a single-electron interacting with constant static potential sources. To our knowledge this is a new result. As an example, suppose that $\mu=1$ (that is $E=\hbar\om$) we find using \eqref{E} and \eqref{dy} that the maser capacitance volume should be $\V=244 \tau_p^2$ if minimum noise is to be achieved. With $\tau_p=1\;\mu$s and $d=0.44\;\mu$m as in \S \ref{static}, the capacitance size $\sqrt A=23$ mm.

\section{Conclusion}\label{conclusion}

A quiet (or sub-Poissonian) oscillator generates a number of dissipation events whose variance is less than the mean. We considered in the present paper oscillators that should exhibit that property, in particular a battery-driven laser involving a single electron permanently interacting with the field. In that case it is unnecessary to quantize the optical field, that is, the theory is semi-classical, yet exact, aside from the approximation made in every above-threshold laser theory that the fluctuations considered are small and slow and the power is small.

We found that if a single-electron laser is driven by a constant-potential battery the detected noise is 7/8 of the shot-noise level, and is therefore sub-Poissonian, for appropriate values of the parameters. This is apparently a new result. Our calculations are related to resonance-fluorescence treatments but with a different physical interpretation. The theory was presented for a single electron. Generalization to many electrons is straightforward if the electrons are not coupled directly to one another through the Coulomb interaction or the Pauli exclusion principle. As is the case for resonance fluorescence with many atoms, anti-bunching (sub-Poissonian radiation) tends to be suppressed. Since sizable amounts of power require a large number of electrons, the battery-driven one-electron laser described above is not a practical way of generating quiet radiation. The latter, at significant power levels, requires non-fluctuating pumps. The case of a non-fluctuating pump ($J$=constant but $U$ or $\gamma$ fluctuating), may be treated by the method explained in this paper.

Lasers involving complicated circuits may also be treated in that manner, distinguishing conservative (loss-less, gain-less) components, treated by the methods of classical electromagnetism, and elements with gain or loss, that should be treated as was done here for a single electron. The latter involve noise sources. These noise sources are at the shot-noise level if the element is linear and the electrons reside most of the time either in the lower state (loss) or in the upper state (gain). But departures from the shot-noise level occur when the element response is non-linear.

\bibliography{quietlasers}

\begin{thebibliography}{12}
\bibitem[{not({\natexlab{a}})}]{note1}
\bibinfo{note}{Light beams carry information if they are modulated in amplitude
  or phase. Small modulations may be obtained from the present theory by
  ignoring the noise sources, but they are not discussed explicitly for the
  sake of brevity.}
\bibitem[{not({\natexlab{b}})}]{note2}
\bibinfo{note}{The precise value of the battery potential $U>\hbar\om/e$ could
  be determined from a detailed analysis of the electron tunneling effect, but
  this needs not be done here. Under the ideal conditions presently assumed,
  the efficiency, that is, the ratio of the detected power and the power
  supplied by the battery, is unity if the power transfer is small. As we
  increase the power transfer, $U$ slightly exceeds $\hbar\om/e$, and some
  irreversible loss occurs. The situation is analogous to that of a system
  consisting of a heat pump raising some amount of heat from a cold to a hot
  reservoir, followed by a heat engine enabling us to recover, ideally, all the
  energy initially supplied to the heat pump. If the power is large, however,
  one must allow some temperature differences to exist between contacted
  bodies, the system is no-longer reversible, and the over-all efficiency
  decreases somewhat.}
\end{thebibliography}

\end{document}